\documentclass[twocolumn,aps,prd,floatfix,preprintnumbers,a4paper,nofootinbib,superscriptaddress]{revtex4-1}

\usepackage{color}
\usepackage{graphicx}													
\usepackage{amssymb}
\usepackage{amsmath}
\usepackage{textcomp}
\usepackage{commath}
\usepackage{bm}
\usepackage{booktabs}
\usepackage{placeins}
\usepackage{xfrac}
\usepackage{gensymb}
\usepackage{adjustbox}

\begin{document}

\newcommand{\Cardiff}{School of Physics and Astronomy, Cardiff University, Queens Buildings, Cardiff CF24 3AA, UK}

\title{Inferring black-hole orbital dynamics from numerical-relativity gravitational waveforms}

\author{Eleanor Hamilton}
\affiliation{\Cardiff}
\author{Mark Hannam}
\affiliation{\Cardiff}

\begin{abstract}
Binary-black-hole dynamics cannot be related to the resulting gravitational-wave signal by a constant retarded time. 
This is due to the non-trivial dynamical spacetime curvature between the source and the signal. In a numerical-relativity
simulation there is also some ambiguity in the black-hole dynamics, which depend on the gauge (coordinate) choices used
in the numerical solution of Einstein's equations. It has been shown previously that a good approximation to the direction of 
the binary's time-dependent orbital angular momentum $\mathbf{\hat{L}}(t)$ can be calculated from the gravitational-wave signal.
This is done by calculating the direction that maximises the quadrupolar $(\ell=2,|m|=2)$ emission. The direction 
depends on whether we use the Weyl scalar $\psi_4$ or the gravitational-wave strain $h$, but these directions 
are nonetheless invariant for a given binary configuration. We treat the $\psi_4$-based direction as a proxy to $\mathbf{\hat{L}}(t)$.
We investigate how well the the binary's orbital phase, $\phi_{\rm orb}(t)$, can also be estimated from the signal. 
For this purpose we define a quantity $\Phi(t)$ that agrees well with $\phi_{\rm orb}(t)$.
One application is to
studies that involve injections of numerical-relativity waveforms into gravitational-wave detector data. 
\end{abstract}

\date{\today}

\maketitle

\section{Introduction}
\label{sec:intro}

In recent years the LIGO and Virgo detectors~\cite{TheLIGOScientific:2014jea,TheLIGOScientific:2016agk,TheVirgo:2014hva} 
have made the first observations of binary-black-hole (BBH) systems, 
through measurements of their gravitational-wave (GW) emission~\cite{Abbott:2016blz,Abbott:2016nmj,Abbott:2017vtc,Abbott:2017oio,Abbott:2017gyy}.
The properties of the black holes can be measured by comparing the signal against theoretical GW models~\cite{TheLIGOScientific:2016wfe,TheLIGOScientific:2016pea}, 
which are informed in part by numerical-relativity (NR) solutions of Einstein's equations for the last orbits and 
merger of two black holes (see, e.g., the review Ref.~\cite{Hannam:2013pra}). NR waveforms have also been used to assess the systematic errors of the GW 
measurements~\cite{Abbott:2016wiq}. To use NR waveforms as proxy signals one must specify the binary's 
orientation and orbital phase at a particular time or signal frequency. There is an inherent ambiguity in doing this,
because the binary's dynamics cannot be directly related to the waveform. The purpose of this work is to 
define an effective binary orientation and phase, which can be calculated directly from the waveform, and compare it against 
the coordinate dynamics in NR simulations.

The general theory of relativity predicts gravitational waves that travel at the speed of light, $c$. 
(Throughout this paper we will adopt geometric units, $G = c = 1$.)
In principle, we 
can relate the dynamics of two orbiting black holes to a GW signal a distance $d$ away, through a retarded time,
$t_{\rm GW} = d/c$. This is possible in a post-Newtonian (PN) calculation~\cite{Blanchet:2013haa}, where the signal can 
be calculated explicitly from point-particle
dynamics. An equivalent identification has not been rigorously defined for solutions of the full nonlinear 
Einstein equations, which are calculated numerically. The proper distance from the source to the observer is not a
well-defined concept. We lack unique definitions of mass, angular momentum and centre-of-mass in general 
relativity~\cite{Szabados:2009eka}; in a numerical simulation the binary dynamics depend non-trivially on the 
gauge (coordinate) conditions used in the evolution of Einstein's equations; proper distances depend on the 
dynamical curvature across the intervening spacetime; and gravitational waves are only rigorously 
defined at null infinity. In practice, these formal ambiguities lead to negligible uncertainties in GW signal modelling
and source measurements; see, for example, Sec.~IV.B of Ref.~\cite{Santamaria:2010yb} in the case of waveform 
modelling, and Ref.~\cite{Boyle:2009vi} for a discussion of retarded times in NR simulations. 

The situation is different when we wish to use NR waveforms as proxy signals. A binary configuration is specified by
the black-hole masses and spin magnitudes, but also by the binary orientation, orbital phase, and spin directions at a 
particular time or frequency during the binary's inspiral. Now we must relate the dynamics to the signal. Given the above, 
we are forced to make approximations. One way to do this is to define an approximate retarded time. Another is to note
that during the inspiral the frequency of the dominant signal harmonic is, to a good approximation, twice the orbital frequency,
and to map the dynamics at each orbital frequency to the corresponding signal frequency. A similar mapping can be
made using the orbital and signal phases, although the two approaches will not give identical results, as we 
discuss in Sec.~\ref{sec: opt ts}.

In this work, we take a different approach. We define a binary orientation and phase with respect to the GW
signal only. The starting point is the earlier work in Ref.~\cite{schmidt2011tracking}, which proposed studying the
direction of maximum GW emission, which was called the direction of ``quadrupole alignment'' (QA). 
The results in Ref.~\cite{schmidt2011tracking} suggested that the QA direction may track the 
direction of $\mathbf{\hat{L}}$. 
If $\mathbf{\hat{L}}$ is calculated using a PN approximation, then the leading-order (Newtonian) contribution
is the normal to the orbital plane, which exhibits nutation, but when all known PN terms are included, the 
full $\mathbf{\hat{L}}$ precesses smoothly. In the NR example studied in Ref.~\cite{schmidt2011tracking}, the
QA direction precessed smoothly without nutation and agreed well with the (appropriately time-shifted) direction 
of $\mathbf{\hat{L}}$. This lead the authors to suggest that the QA direction may track the orbital angular momentum, 
rather than the orbital-plane direction. More recent work has shown that this direction varies between different radiation
frames, and also depends on whether the direction is calculated using the GW strain $h$, the Bondi news (the first time
derivative of $h$), or the Weyl scalar $\psi_4$ (the second time derivative)~\cite{Lousto:2013wta,Boyle:2014ioa}. 
Nonetheless, in general these differences are small, and any given choice of the QA direction provides us with an ideal 
means to define a proxy to the binary orientation with respect to the GW signal alone. Since the GW signal is the only 
invariant observable we have access to, this orientation provides a robust measure to identify and compare simulations. 

The first QA definitions~\cite{schmidt2011tracking,o2011efficient} specified only the two Euler angles needed to transform into a
frame that tracks the precession. A third Euler angle is also needed to uniquely specify the phase (up to an overall constant). A method
to calculate the third angle is given in Ref.~\cite{boyle2011geometric}, completing the definition of a co-precessing frame. In this work
we use that procedure to define a proxy orbital phase, $\Phi$, from the GW signal, which in turn allows us to define a proxy orbital 
separation unit vector $\mathbf{\hat{n}}$, which we compare with those quantities calculated directly from the orbital dynamics. 
Once again, we show that this does not provide an exact mapping to the phase calculated directly 
from the dynamics, even if time shifts and gauge effects could be removed; but $\Phi$ does serve as a phase that is in principle gauge invariant and uniquely defined. 

To connect our work to the practical problem of constructing proxy GW signals from NR waveforms, we describe our work
and results using the notation and conventions of the NR Injection Infrastructure~\cite{schmidt2017numerical}, which
provides a consistent way to go from waveforms produced using a variety of NR codes to waveforms that are suitable for 
injections as a ``discrete'' waveform approximant for use with the LIGO Algorithm Library (LAL). The LAL framework requires injected 
waveforms to be in a frame that describes the wave propagation from the source to GW detectors on Earth. The NR Injection Infrastructure 
rotates the waveforms into this format. These rotations require the unit orbital angular momentum of the binary, $\mathbf{\hat{L}}$, 
and the unit separation vector of the two black holes, $\mathbf{\hat{n}}$. The unit separation vector $\mathbf{\hat{n}}$ can be constructed
from the normal to the orbital plane and the orbital phase; our approach will be to define $\mathbf{\hat{n}}$ from $\mathbf{\hat{L}}$
and $\Phi$. 
These quantities are currently calculated using the dynamics information provided by a simulation. To relate these 
dynamical data to the GW signal, one either uses an estimate of the retarded time $t_{\text{GW}}$ (provided along with the NR waveform,
and corresponding to Format 1 in Ref.~\cite{schmidt2017numerical}), or maps the orbital frequency $\Omega(t)$ to the signal frequency (Formats
2 and 3). The method we propose is equivalent to mapping the orbital \emph{phase} to that of the signal, 
and without any of the gauge ambiguities of the black-hole coordinate dynamics. 

The paper is organised as follows. In Sec.~\ref{sec: conventions} we describe the rotations performed by the NR Injection Infrastructure. 
In Sec.~\ref{sec: Landn} we summarise the procedure to find the unit orbital angular momentum, which is described in more detail in 
Refs.~\cite{schmidt2011tracking, o2011efficient, boyle2011geometric}, and describe how to also find the coprecessing phase and the unit separation vector from the waveform. 
Section \ref{sec:ambiguities} describes the various coordinate ambiguities associated with these calculations.
In Sec.~\ref{sec:comparisons} we compare $\mathbf{\hat{L}}$, $\Phi$, and $\mathbf{\hat{n}}$, which have been calculated from the waveform, 
with those found from the dynamics. We also discuss how the different choices of time shift affect this comparison and show why it is important 
to ensure a consistent choice is used.

\section{Frame conventions}
\label{sec: conventions} 

\begin{figure}[htbp]
   \centering
   \includegraphics[scale=0.5]{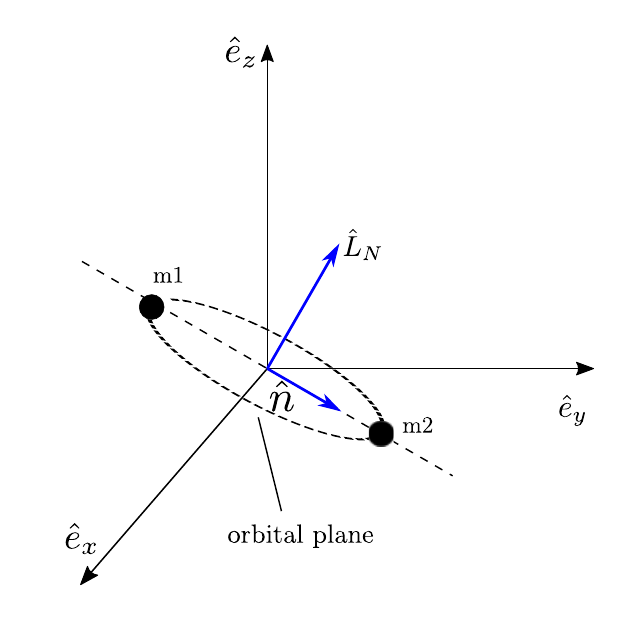} 
   \caption{The binary properties in the NR simulation frame (black) at a time $t_{\rm ref}$. The binary is then rotated to the LAL source frame (blue) where $\mathbf{\hat{z}}$ is parallel to the (Newtonian) orbital angular momentum $\mathbf{\hat{L}_N}$ at time $t_{\rm ref}$ and $\mathbf{\hat{x}}$ is aligned along $\mathbf{\hat{n}}$.
   }
   \label{fig: frames}
\end{figure}

In this section we summarise three coordinate systems used to specify GW signals. We follow the conventions and notations used in the Numerical Relativity Injection 
Infrastructure~\cite{schmidt2017numerical}. GW signals are represented by the gravitational-wave strain, which corresponds to the metric perturbation $h_{ij}^{\text{TT}}$. Numerical simulations calculate the Weyl scalar $\psi_4$, from which $h_{ij}^{\text{TT}}$ can be found by integrating twice with respect to time~\cite{Reisswig:2010di}. 
In numerical simulations this perturbation is extracted far from the orbiting black holes, where the spacetime is approximately flat. This region of spacetime is known as the \textit{wave zone} \cite{thorne1980multipole}. The waves are extracted at a retarded time $t_{\text{GW}}$. In the wave zone, a Cartesian co-ordinate system $\left(\mathbf{\hat{e}_{x}},\mathbf{\hat{e}_{y}},\mathbf{\hat{e}_{z}}\right)$ is used. 
This co-ordinate system can be related to polar co-ordinates $\left(\mathbf{\hat{e}_{r}},\mathbf{\hat{e}_{\theta}},\mathbf{\hat{e}_{\phi}}\right)$. 
The strain can then be decomposed into modes in a basis of spin-weighted spherical harmonics, $^{-2}Y_{\ell m}$, and is written as 
\begin{align} 
   h^{\text{NR}}\left(t_{\text{GW}};\theta,\phi\right)  = & h_{+}^{\text{NR}} - i h_{\times}^{\text{NR}} \\
    = & \sum^{\infty}_{\ell = 2} \sum^{\ell}_{m = -\ell} H_{\ell m}\left(t_{\text{GW}}\right) ^{-2}Y_{\ell m}\left(\theta,\phi\right),
\end{align}
where the extracted GW modes can be expressed as
\begin{align} 
   H_{\ell m}\left(t_{\text{GW}}\right) = {}& A_{\ell m}\left(t_{\text{GW}}\right) e^{- i\Phi_{\ell m}\left(t_{\text{GW}}\right)}.
\end{align}
We adopt the convention that for a binary orbiting counter-clockwise in the plane defined by $\mathbf{\hat{e}_{x}} \times \mathbf{\hat{e}_{y}}$, 
$\Phi_{22}\left(t_{\text{GW}}\right)$ is a monotonically increasing function.

Once the GW has been extracted and decomposed as described above it needs to be prepared for injection. This involves transforming the waveform from the frame in which it has been generated (the \textit{NR simulation frame}) into the frame in which the binary is viewed from Earth. This is done in two stages. First, the waveform is rotated into a frame defined by certain properties of the binary at a given reference time. The choice of this frame is arbitrary but must be consistent between injections. A set of conventions in defining this frame, known as the \textit{LAL source frame} ~\cite{LALsimulationframes, schmidt2017numerical}, are therefore chosen. These conventions are described below. In this frame, waveforms generated by a particular binary should be the same regardless of the code used to generate them or the choice of coordinate system in the original simulation. From this intermediate frame, the waveform is then rotated into the final frame, the \textit{wave frame}, defined by the relationship between the binary and the observer.

In the NR simulation frame one can define the separation vector of the two black holes as the direction from body 2 to body 1 (where body 1 is the heavier object) given by
\begin{align} \label{eqn: nhat}
   \mathbf{n} = {}& \mathbf{r_{1}} - \mathbf{r_{2}},
\end{align}
where $\mathbf{r_{i}}$ is the position of the centre of the $i$th body. The Newtonian orbital angular momentum of the binary can be defined as
\begin{align} \label{eqn: lhat}
   \mathbf{L_{N}} = {}& \mathbf{L_{1}} + \mathbf{L_{2}} = \sum_{i=1}^{2} m_{i} \left(\mathbf{r_{i}} \times \mathbf{v_{i}}\right),
\end{align}
where $m_{i}$ is the mass and $\mathbf{v_{i}}$ the velocity of the $i$th object. In moving-puncture codes, $\mathbf{r_{i}}$ will be the puncture positions, and in excision codes they
will be the coordinate centres of the apparent horizons. 

The LAL source frame is defined as the frame where the coordinate axes satisfy the following equalities
\begin{align} \label{eqn: LALx}
   \mathbf{\hat{x}} \overset{\mathrm{ref}}{=} {}& \mathbf{\hat{n}} \\ 
   \mathbf{\hat{y}} \overset{\mathrm{ref}}{=} {}& \mathbf{\hat{L}_{N}} \times \mathbf{\hat{n}} \\
   \mathbf{\hat{z}} \overset{\mathrm{ref}}{=} {}& \mathbf{\hat{L}_{N}} \label{eqn: LALz}
\end{align}
at a reference epoch defined either by a reference time $t_{\rm ref}$ or a reference orbital frequency $\Omega^{\rm orb}_{\rm ref}$ where $\Omega^{\rm orb}\left(t_{\rm ref}\right) = \Omega^{\rm orb}_{\rm ref}$. Choosing a different reference epoch will in general produce a different source frame. 

Finally, the waveform is rotated into the wave frame. In this frame the $\mathbf{\hat{Z}}$ axis points towards the observer along the line of sight while 
the $\mathbf{\hat{X}}$ and $\mathbf{\hat{Y}}$ vectors are orthogonal to the line of sight. The intersection of the orbital plane with the $XY$ axis is referred 
to as the line of ascending node. The transformation to the wave frame is given in Ref~\cite{schmidt2017numerical}; for the remainder of this paper 
we will work in either the inertial NR simulation frame or the co-precessing (quadrupole-aligned) frame, as described in Sec.~\ref{sec: Landn}.

Ambiguities in this procedure arise from the NR simulation data. 
The NR simulation frame is the coordinate system in which the numerical simulation was performed. The physical interpretation of the coordinates
in the NR simulation frame depends on the coordinates of the initial data, and on the gauge conditions used during the numerical evolution. If simulations with two different codes, using
different initial-data constructions and different gauge conditions, are used to simulate the same physical system, then in principle we expect the asymptotic gravitational-wave signals
to be the same, but the black-hole dynamics in the respective NR simulation frames may not be. We aim to circumvent these ambiguities in the method that we propose in the following sections. 

\section{Determining $\mathbf{\hat{L}_{N}}$ and $\mathbf{\hat{n}}$}
\label{sec: Landn}

Currently the NR Injection Infrastructure calculates $\mathbf{\hat{n}}$ and $\mathbf{\hat{L}_{N}}$ using Eqs.~(\ref{eqn: nhat}) and (\ref{eqn: lhat}) respectively (and then normalising). 
The positions and velocities of the black holes required to calculate these quantities come from the dynamics of the binary.   This information, along with the spins of the black holes, 
forms part of the metadata provided with each NR waveform. There are several sources of ambiguity in the resulting choice of LAL source frame (as defined via 
Eqs.~(\ref{eqn: LALx})--(\ref{eqn: LALz})). One is the gauge dependence of the coordinate dynamics
and spin measurements. (Broadly speaking, codes that use variants of the generalized-harmonic evolution system~\cite{Pretorius:2004jg,Pretorius:2005gq}, like 
SpEC~\cite{Scheel:2008rj,Szilagyi:2014fna}, use harmonic-like coordinates~\cite{Lindblom:2005qh}, while moving-puncture codes~\cite{Baker:2005vv,Campanelli:2005dd} use ADMTT-like 
coordinates~\cite{Jaranowski:1997ky}. For an example of one comparison between these coordinates,
see Appendix~D  of Ref.~\cite{Brugmann:2007zj}.)  The black-hole dynamics information can be mapped to the waveform using either a retarded time
(Format 1 in the NR Injection Infrastructure), or relating the GW frequency with the orbital frequency (Formats 2 and 3). If the retarded time is used, then
a further ambuguity arises from the definition of retarded time $t_{\text{GW}}$ used by a particular NR group. 
These ambiguities could be resolved by finding $\mathbf{\hat{n}}$ and $\mathbf{\hat{L}}$ from the  waveform. The NR waveform includes some error due to extraction at a finite coordinate radius, or due to approximate 
extrapolation to infinity, but in general exhibits far less gauge variation than the dynamics. We now describe a procedure to do this. 

\subsection{Determining $\mathbf{\hat{L}}$ from the waveform}\label{sec: L}

It has already been shown that the direction of the orbital angular momentum of a binary can be found from the waveform of the emitted GWs using a variety of methods ~\cite{schmidt2011tracking,o2011efficient}. 
Ref~\cite{o2011efficient} describes how this quantity can be found from the dominant principal axis of the quadrupolar part of the radiation axis. We use the Quadrupole Alignment procedure 
described in Ref~\cite{schmidt2011tracking}; the two methods can be shown to be equivalent~\cite{boyle2011geometric}. This procedure finds the frame in which $|\psi_{4,22}|^2 + |\psi_{4,2-2}|^2$ is 
maximised. In this frame $\mathbf{\hat{z}} \parallel \mathbf{\hat{L}}$. This transformation requires two angles ($\alpha$ and $\beta$), which define the rotation into a coprecessing frame, i.e.,  a frame that 
precesses along with the binary. In order to uniquely define this frame (up to an overall constant rotation, corresponding to a constant phase shift in the waveform in the coprecessing frame) we apply the minimum rotation condition \cite{boyle2011geometric}, which gives the third Euler angle as 
\begin{align} 
   \dot{\gamma} = {}& -\dot{\alpha}\cos{\beta}. 
\end{align}
This angle is determined up to an integration constant, which corresponds to a constant rotation.
A time-dependent rotation \textbf{R}$\left(\gamma\beta\alpha\right)$ can then be performed between the inertial frame in which the simulation was performed and the coprecessing frame using the three Euler angles $\left(\alpha,\beta,\gamma\right)$. Using the z-y-z convention, the $\psi_{4,\ell m}$ modes obey the transformation law
\begin{align} 
   \psi_{4,\ell m}^{QA} = {}& \sum_{m'=-\ell}^{\ell} e^{im'\gamma}d^{\ell}_{m'm}\left(-\beta\right)e^{im\alpha}\psi_{4,\ell m'}
\label{eqn: transformation law}
\end{align}
where $\psi_{4,\ell m'}$ are the modes in the NR simulation frame and $\psi_{4,\ell m}^{QA}$ are the modes in the coprecessing (quadrupole-alinged) frame. $d^{\ell}_{m'm}$ are the Wigner $d$-matrices~\cite{wigner2012group, goldberg1967spin}.

The coprecessing frame rotates with the orbital angular momentum in order to ensure $\mathbf{\hat{L}}$ remains parallel to the $z$-axis at all times. Since $\mathbf{\hat{L}}$ is approximately perpendicular to the orbital plane, the orbital plane remains approximately in the $xy$-plane in the coprecessing frame.

\subsection{Determining orbital phase and $\mathbf{\hat{n}}$ from the waveform}\label{sec: nhat}

During the early inspiral of a non-precessing binary, the orbital phase of the binary can be found from the phase of the waveform, using 
\begin{align} 
   \omega_0^{\ell m} = {}& m \od{}{t}\phi_{\rm orb}\left(t_0\right),
\label{QNMeqn}
\end{align}
where $\omega_0^{\ell m}$ is the angular frequency of $\psi_{4,\ell m}$ and $\phi_{\rm orb}$ is the orbital phase of the binary in the orbital plane. 
PN corrections to this relation are small~\cite{Blanchet:2013haa}, the differences between the phases of $h$ and $\psi_4$ are also small~\cite{Boyle:2007ft},
and this approximations holds to high accuracy even up until a few orbits before 
merger (see, e.g., Fig.~7 of Ref.~\cite{Buonanno:2006ui}), Consequently, the orbital phase of the binary is half that of the phase of the $\psi_{4,22}$ mode. 
The phase of a $\psi_{4, \ell m}$ mode, $\Phi^{\psi_4}_{\ell m}$, is the unwrapped argument of the complex time series $\psi_{4, \ell m}$ given by
\begin{align} 
   \psi_{4,\ell m} = {}& A^{\psi_4}_{\ell m} e^{-i \Phi^{\psi_4}_{\ell m}}.
\end{align}
As stated above, the phase of the $\left(2,2\right)$ mode is a monotonically increasing function. Therefore, once the orbital phase has been calculated the unit separation vector is given by 
\begin{align} \label{n}
   \mathbf{\hat{n}^{\text{QA}}} = {}& \begin{pmatrix} \cos\phi_{\rm orb} \\ -\sin\phi_{\rm orb} \\ 0 \end{pmatrix} \approx \begin{pmatrix} \cos\left[\frac{1}{2}\left(\Phi^{\psi_4, \rm{QA}}_{22} + \Phi_0\right)\right] \\ -\sin\left[\frac{1}{2}\left(\Phi^{\psi_4, \rm{QA}}_{22} + \Phi_0\right)\right] \\ 0 \end{pmatrix},
\end{align}
where $\Phi_0$ is the orbital phase offset, which depends on the conventions used in the NR code used to produce the simulation. It is 0 if the phase of $\psi_4$ is $0\mod 2\pi$ when the black holes are on the $x$-axis, and $\pm\pi$ if this occurs when they are on the $y$-axis. For the remainder of this paper, we will define the orbital phase, as estimated from the waveform, as $\Phi = (\Phi_{22}^{\psi_4,QA} + \Phi_0)/2$.

For a precessing binary, a similar procedure can be performed by rotating the waveform into the co-precessing frame described in section \ref{sec: L}. 

The unit separation vector can then be found as for a non-precessing waveform. It then needs to be rotated back into the NR simulation frame using the angles $\alpha$, $\beta$ and $\gamma$ found above. Since these angles were defined using the z-y-z convention, the rotations required to rotate a vector from the Quadrupole Aligned frame to the NR simulation frame are 
\begin{itemize}
\item rotate by $\gamma$ about the z-axis
\item \underline{then} rotate by $\beta$ about the y-axis
\item \underline{then} rotate by $\alpha$ about the z-axis.
\end{itemize}
This is given by
\begin{widetext}
\begin{align} 
      & \begin{pmatrix} 
         \cos\alpha & -\sin\alpha & 0 \\
         \sin\alpha & \cos\alpha & 0 \\
         0 & 0 & 1 \\
      \end{pmatrix} 
      \begin{pmatrix} 
            \cos\beta & 0 & -\sin\beta \\
            0 & 1 & 0 \\
            \sin\beta & 0 & \cos\beta \\
      \end{pmatrix}
      \begin{pmatrix}
         \cos\gamma & -\sin\gamma & 0 \\
         \sin\gamma & \cos\gamma & 0 \\
         0 & 0 & 1 \\
      \end{pmatrix}
      \begin{pmatrix} 
            \hat{n}_{x}^{\text{QA}} \\ \hat{n}_{y}^{\text{QA}} \\ \hat{n}_{z}^{\text{QA}}
      \end{pmatrix} \\
      = {}& \begin{pmatrix} 
            \cos\alpha\left(\cos\beta \left(\cos\gamma \hat{n}_{x}^{\text{QA}} - \sin\gamma \hat{n}_{y}^{\text{QA}}\right) - \sin\beta \hat{n}_{z}^{\text{QA}}\right) - \sin\alpha \left(\sin\gamma \hat{n}_{x}^{\text{QA}} + \cos\gamma \hat{n}_{y}^{\text{QA}}\right)\\
            \sin\alpha\left(\cos\beta \left(\cos\gamma \hat{n}_{x}^{\text{QA}} - \sin\gamma \hat{n}_{y}^{\text{QA}}\right) - \sin\beta \hat{n}_{z}^{\text{QA}}\right) + \cos\alpha \left(\sin\gamma \hat{n}_{x}^{\text{QA}} + \cos\gamma \hat{n}_{y}^{\text{QA}}\right) \\
            \sin\beta \left(\cos\gamma \hat{n}_{x}^{\text{QA}} - \sin\gamma \hat{n}_{y}^{\text{QA}}\right) + \cos\beta \hat{n}_{z}^{\text{QA}}
         \end{pmatrix}. \nonumber
\end{align}
\end{widetext}

Since the waveform is rotated into the co-precessing frame by the three Euler angles, the orientation of $\mathbf{\hat{n}^{\text{QA}}}$ is determined up to a constant phase based on the choice of integration constant when calculating $\gamma$. However, when rotating $\mathbf{\hat{n}^{\text{QA}}}$  into the NR simulation frame, the rotation by $\gamma$ removes this ambiguity meaning 
$\mathbf{\hat{n}}$ is uniquely determined in the NR simulation frame regardless of the choice of integration constant.

\subsection{Determining the coprecessing orbital phase}
\label{sec: orbphase}

Alternatively, one can rotate the unit separation vector $\mathbf{\hat{n}_{d}}$ (calculated from the positions of the black holes) into the coprecessing frame. This involves performing the above rotations in the reverse order using the Euler angles calculated from the Newtonian orbital angular momentum. The coprecessing orbital phase can then easily be calculated.

Since the Euler angle $\gamma$ is found using integration a constant is introduced into the coprecessing phases. We determined this constant using the fact that $\arccos\left(\mathbf{\hat{n}_{w}}\cdot\mathbf{\hat{n}_{d}}\right) = \Phi - \phi_{\rm orb}$.

\subsection{Code Conventions}
\label{sec:conventions}

Several convention choices enter into the calculation of $\psi_4$. These determine the relationship between the phase of $\psi_4$ and the orbital phase of the binary, i.e., they determine 
the orbital phase offset $\Phi_0$ given in Eq.~(\ref{n}). The three relevant choices here are the sign convention in the definition of the Riemann and Weyl tensors, the definition of $\psi_4$ itself and the choice of origin of the azimuthal angle $\varphi$ of the spherical co-ordinates. The first two of these differences introduce an ambiguity in the definition of $\psi_4$ of $\psi_4 \longrightarrow e^{i \psi_0}\psi_4$. The third introduces the ambiguity $\psi_{4, \ell m} \longrightarrow e^{i m \varphi_0}\psi_{4, \ell m}$~\cite{bustillo2015comparison}

An example of the effect of different choices in these conventions is the difference in the phase of $\psi_4$ calculated by identical simulations produced using the 
BAM ~\cite{brugmann2008calibration, husa2008reducing} and SpEC~\cite{Scheel:2008rj,Szilagyi:2014fna} codes.
These have been explained in  Ref.~\cite{bustillo2015comparison}. The two codes use the opposite sign convention in the definition of the Riemann and Weyl tensors. Additionally, a different choice of null tetrad is made when defining $\psi_4$; in the BAM code, $\psi_4$ is defined via $\psi_4 = -C_{\alpha\mu\beta\nu}n^{\mu}n^{\nu}\bar{m}^{\alpha}\bar{m}^{\beta}$ \cite{bustillo2015comparison}, while in the SpEC code $\psi_4$ is defined by $\psi_4 = -C_{\alpha\mu\beta\nu}\ell^{\mu}\ell^{\nu}\bar{m}^{\alpha}\bar{m}^{\beta}$ (see Ref.~\cite{pfeiffer2007reducing} and Sec.~4.3.1 of Ref.~\cite{chu2012numerical}). $\left(\ell^{\mu}, m^{\mu}, \bar{m}^{\mu}, n^{\mu}\right)$ is an appropriate null tetrad where $\ell$ and $n$ are ingoing and outgoing null vectors respectively and $-\ell\cdot n = 1 = m\cdot\bar{m}$. $C_{\alpha\mu\beta\nu}$ is the Weyl tensor. 
These choices produce a phase offset of $\pi$ ($\psi_0 = -1$) between $\psi_4$ calculated by BAM and by SpEC at equivalent points in the waveform for an identical simulation. The choice of the origin of $\varphi$ can differ between simulations. However it seems that on the whole the choices made by BAM and SpEC do not introduce any additional phase offset.

These different choices of conventions mean that for BAM the phase of $\psi_4$ is $(0\mod 2\pi)$ when the two black holes are on the $x$-axis (of the co-precessing frame) whereas for SpEC this happens when the two black holes are on the $y$-axis. Consequently, in order to calculate a value of $\mathbf{\hat{n}}$ which agrees with the dynamics information provided along with a simulation,
\begin{align}
   \mathbf{\hat{n}^{\text{QA}}_{\text{BAM}}} = {}& \begin{pmatrix} \cos\Phi^{\psi_4, \rm{QA}}_{22} \\ -\sin\Phi^{\psi_4, \rm{QA}}_{22} \\ 0 \end{pmatrix},
\end{align}
while 
\begin{align}
   \mathbf{\hat{n}^{\text{QA}}_{\text{SXS}}} = {}& \begin{pmatrix} -\sin\Phi^{\psi_4, \rm{QA}}_{22} \\ -\cos\Phi^{\psi_4, \rm{QA}}_{22} \\ 0 \end{pmatrix}.
\end{align}
The GT-MAYA~\cite{vaishnav2007matched, healy2009zoom, herrmann2007unequal} and RIT~\cite{zlochower2005accurate} codes appear to use the same conventions as the SpEC code. 
These conventions are also used when producing the PN waveforms outlined in Ref.~\cite{Blanchet:2013haa,arun2009higher}.

In general, a consistent convention for $\Phi_0$ must be chosen. A choice of $\Phi_0 = \frac{\pi}{2}$ agrees with the PN convention. This will give a consistent definition of $\mathbf{\hat{n}}$ from the waveform, regardless of the convention choice of the NR code which determines the dynamics of the simulation. The individual code conventions need to be
taken into account only when we wish to compare back to the coordinate dynamics of the original NR simulation.

\section{Coordinate ambiguities} 
\label{sec:ambiguities}

In this section we illustrate two coordinate ambiguities that we referred to earlier. The first is in the definition of the retarded time $t_{\rm GW}$; we
consider typical choices of retarded time that have been used in numerical-relativity studies, and also the retarded times implied by aligning either the 
GW phase or frequency with the corresponding quantity calculated from the dynamics. The second is in the estimate of the orbital-plane orientation. 
We use a PN example to illustrate these ambiguities, in particular differences in the QA direction calculated using $\psi_4$ and $h$. 
The differences in these directions are nonetheless small, as we illustrate with both PN and
NR examples.

\begin{figure*}[htbp]
   \centering
   \includegraphics[width=\textwidth]{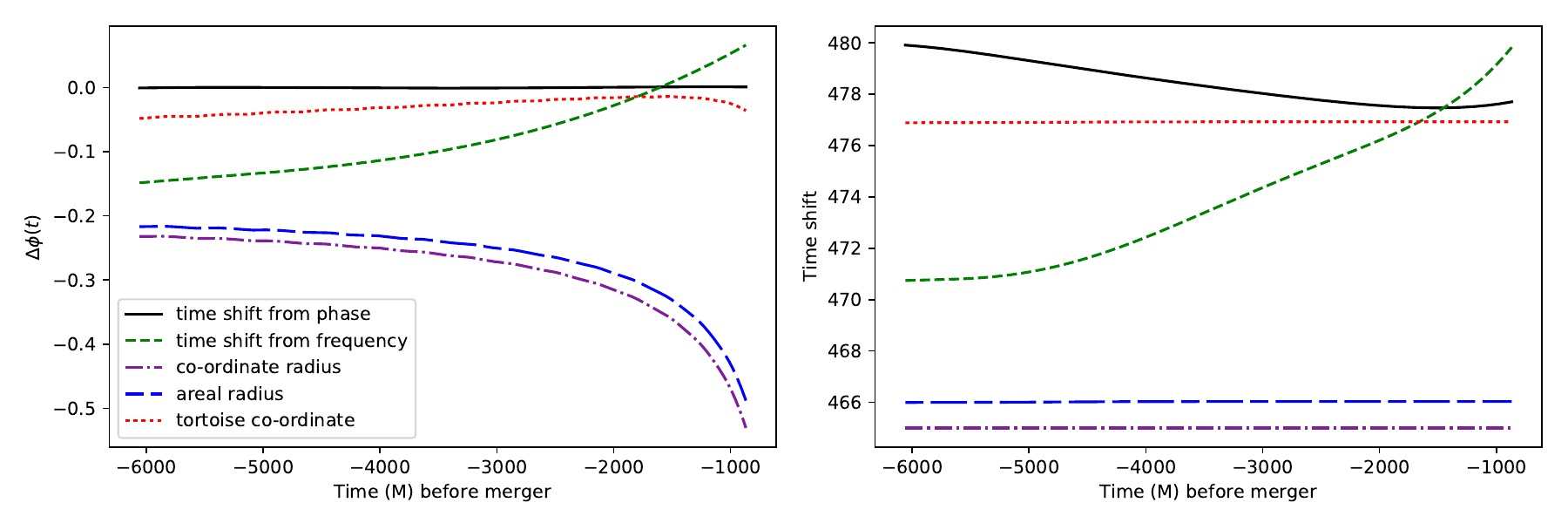} 
   \caption{Simulation SXSBBH0152 ($q = 1$, $M \omega_{22}^{\rm start} = 0.0297$, $\chi_{1} = (0,0,0.6) = \chi_{2}$). The left hand plot shows the difference between the 
   orbital phase estimate from the GW signal, $\Phi$, and the orbital frequency $\phi_{orb}$ for each of the time shifts shown in the right hand plot.
   }
   \label{fig: np diff ts SXS}
\end{figure*}

\begin{figure*}[htbp]
   \centering
   \includegraphics[width=\textwidth]{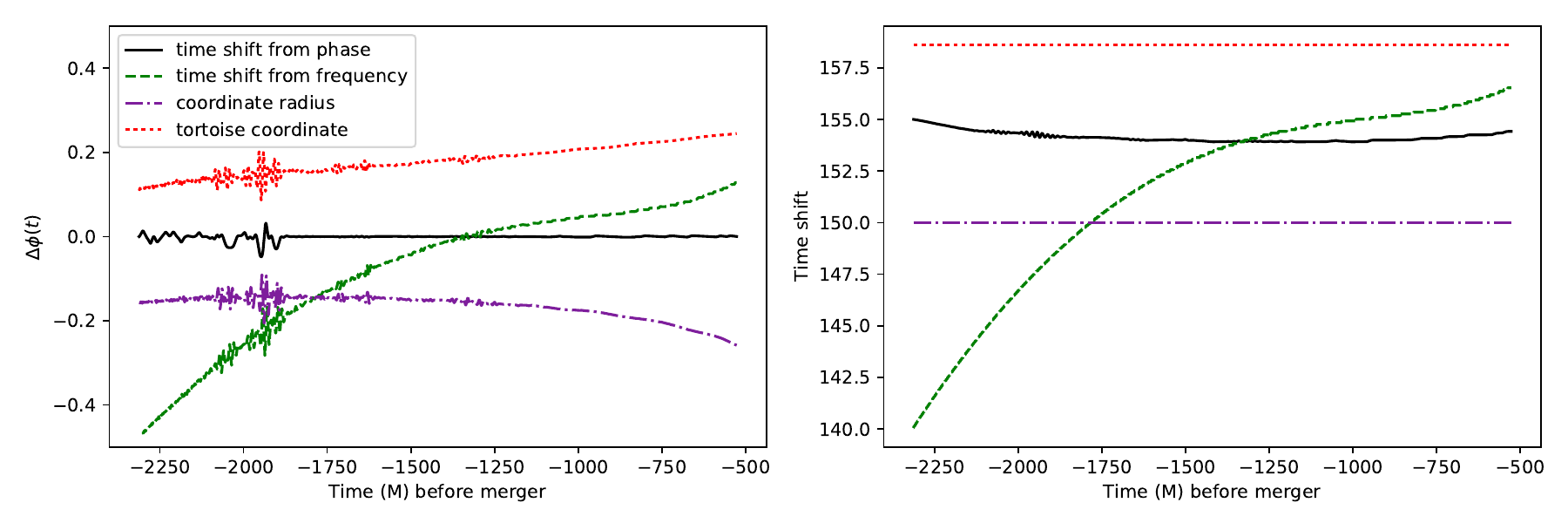} 
   \caption{The same quantities as in Fig.~\ref{fig: np diff ts SXS}, but for the BAM q8 simulation ($q = 8$, $M \omega_{22}^{\rm start} = 0.0625$, $\chi_{1} = (0,0,0.0105)$, $\chi_{2} = (0,0,0.672))$. 
   }
   \label{fig: np diff ts BAM}
\end{figure*}

\subsection{Retarded time}
\label{sec: opt ts}

As mentioned above, there is an ambiguity when relating information about the binary dynamics calculated at the source of the simulation to waveform information extracted at some finite coordinate distance from the source. Different groups use different conventions to define the relationship between the time at the source $t$ and the retarded time $t_{\text{GW}}$. The two methods most commonly used are (i) to treat the spacetime as if it were flat and (ii) to assume the propagation time is given by the tortoise coordinate, as in, for example, Ref.~\cite{boyle2009extrapolating}. The two choices of retarded time can be summarised as \begin{eqnarray}
\text{(i) } t_{\text{GW}} & = & t + R_{ex}, \label{eqn: coord time}  \\ 
\text{(ii) } t_{\text{GW}} & = & t + R_{ex} + 2M\ln\left|\frac{R_{ex}}{2M} -1\right| \label{eqn: tortoise}
\end{eqnarray}
where $M$ is the initial total mass of the system and $R_{ex}$ is the coordinate radius at which the GW signal was extracted from the NR simulation.

These different conventions mean the metadata provided with the waveforms used in the NR Injection Infrastructure are not defined in a consistent manner. The method described in 
Sec.~\ref{sec: Landn} to find $\mathbf{\hat{L}}$ and $\mathbf{\hat{n}}$ removes this ambiguity and provides a consistent way of defining $\mathbf{\hat{L}}$ and $\mathbf{\hat{n}}$ for all waveforms. 
This method is equivalent to using the time shift that aligns the phase of the waveform with the orbital phase at each time step, in the coprecessing frame. 
This time-dependent time shift can then be used to also report 
the spins in a consistent manner.

We will illustrate the difference between these choices with two waveforms from non-precessing binaries. One is an equal-mass-binary waveform selected from the SXS catalog of SpEC waveforms~\cite{SXSdocumentation}, and the other is a mass-ratio 1:8 binary simulated with the BAM code. (These are the SXS 152 and BAM q8 configurations listed in Tab.~\ref{tab: sims} in Sec.~\ref{sec:comparisons}.) We denote the orbital phase of the two black holes by $\phi_d(t)$, and the corresponding phase of the gravitational-wave signal by $\Phi$, as described in Sec.~\ref{sec: nhat}. 
For each choice of retarded time $t_{\rm GW}$, we calculate the phase difference $\Delta \phi (t)  =  \Phi(t_{\rm GW}) - \phi_d(t)$. Figs.~\ref{fig: np diff ts SXS} and \ref{fig: np diff ts BAM} show the results for several choices of retarded time. For the SXS waveform, we consider three choices of retarded time: as defined by the coordinate extraction radius, 
by the areal radius of the extraction sphere, $R_{areal} = \sqrt{A/4\pi}$, where $A$ is the proper area of the extraction sphere~\cite{boyle2009extrapolating, SXSdocumentation}, 
and by the tortoise coordinate calculated from the areal radius. (The tortoise coordinate choice was used to produce the Format 1 metadata for SXS waveforms.)
For the coordinate and areal-radius choices, we see that the phase difference can be as large as 0.5\,rad 1000$M$ before merger. 
The phase difference when using the tortoise coordinate is much smaller, but still non-zero. By construction the phase difference is zero when using the method described in 
Sec.~\ref{sec: Landn}, since it is equivalent to aligning the orbital and GW phases at each time. The signal propagation times implied by each choice are shown in the right panel. 
We see that the propagation time varies with the waveform-based choices, but that is not surprising, given the gauge-dependent nature of the coordinate dynamics. 
The areal radius has not been calculated for the BAM waveform, so Fig.~\ref{fig: np diff ts BAM} shows results only for the coordinate extraction radius, and the tortoise coordinate calculated using this value. We again see that the phase difference is smallest when using the tortoise coordinate. The variation in the time shift required to align the phases is comparable between the SXS and BAM waveforms.

Based on the results in Figs.~\ref{fig: np diff ts SXS} and \ref{fig: np diff ts BAM}, we see that the tortoise coordinate provides the best phase alignment between the dynamics and GW signal for both codes. 
We also find that the results based on our procedure give similar agreement. This procedure has the additional advantages that it can be applied agnostically to all NR waveforms, and is based directly on the gauge-invariant GW signal.

\subsection{Orbital plane nutation}
\label{sec:PN}

In precessing configurations the orbital plane exhibits nutation that is not present in the direction of
the full post-Newtonian orbital angular momentum. Ref.~\cite{schmidt2011tracking} showed that the QA direction calculated
from $\psi_4$ also precesses smoothly, suggesting that this method may be a better approximation to the direction of the orbital angular momentum,
 than to the orbital plane. 
We consider a PN example, and illustrate that although this identification does not hold, the QA direction is nonetheless a good approximation to the 
binary orientation. For an NR configuration we also quantify the differences between
the $\psi_4$ estimates of the binary orientation and phase, and those calculated from the orbital dynamics, and show that they are 
small.

\begin{figure}[htbp]
   \centering
   \includegraphics[width=0.5\textwidth]{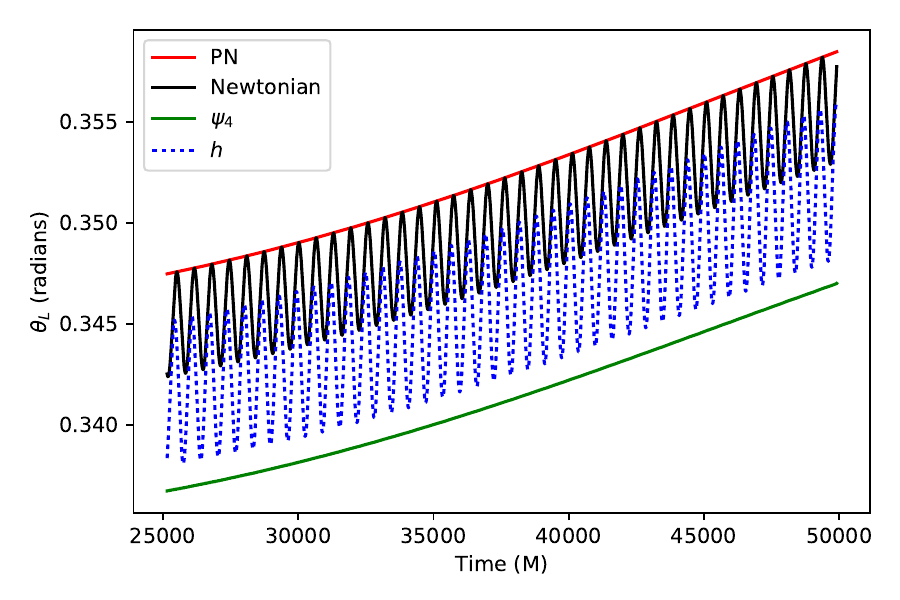} 
   \caption{$\theta_{L}$ calculated from the Newtonian orbital angular momentum (i.e., the normal to the orbital plane), the post-Newtonian
   orbital angular momentum, and from $\psi_4$ and the GW strain. (See text for discussion.)
   }
   \label{fig: L_comparison1}
\end{figure}

\begin{figure}[htbp]
   \centering
   \includegraphics[width=0.5\textwidth]{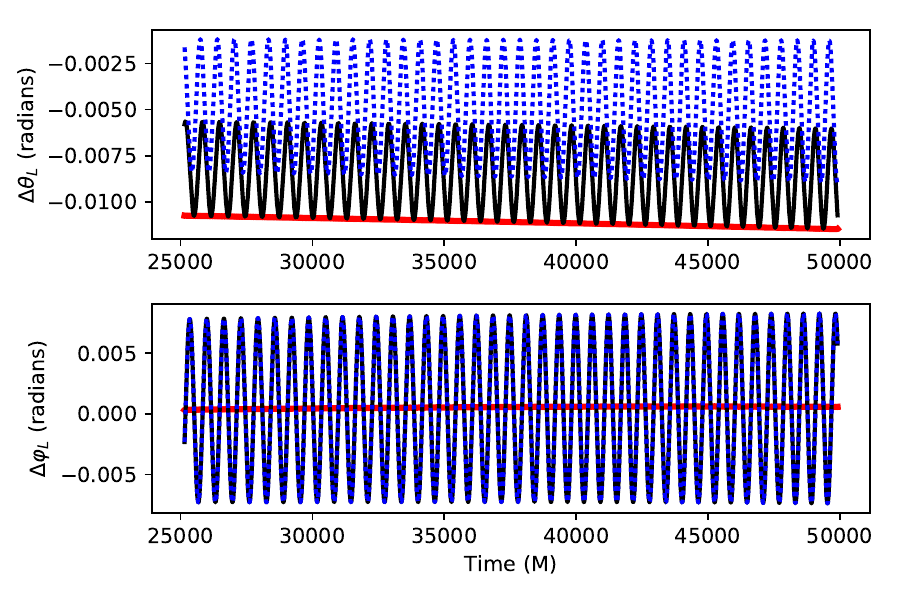} 
   \caption{Difference between the calculation of $\theta_{L}$ and $\varphi_{L}$ from $\psi_4$, and that calculated from $\mathbf{\hat{L}_{N}}$ (black line), 
   $\mathbf{\hat{L}_{PN}}$ (red line) and $\mathbf{\hat{L}_{h}}$ (dashed blue) for a post-Newtonian waveform with $q = 3$, $\chi = 0.75$ on the larger black hole, on average in the orbital plane. 
   }
   \label{fig: L_comparison}
\end{figure}

\begin{figure}[htbp]
   \centering
   \includegraphics[width=0.5\textwidth]{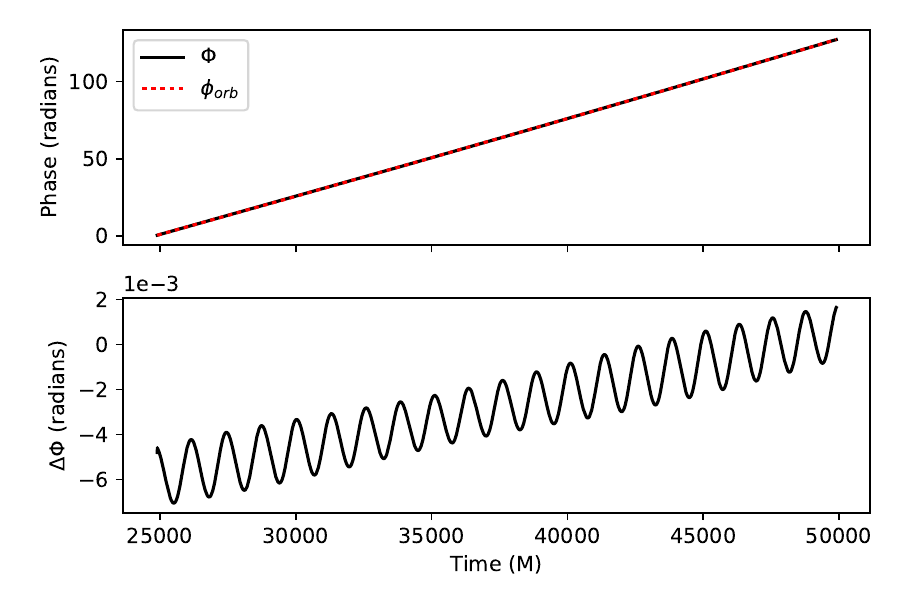} 
   \caption{PN waveform ($q = 3$, $\chi = 0.75$ on the larger black hole, on average in the orbital plane). A comparison of half the coprecessing GW phase $\Phi$ and the orbital phase $\phi_{orb}$.
   }
   \label{fig: PN}
\end{figure}

\begin{figure*}[htbp]
   \centering
   \includegraphics[width=\textwidth]{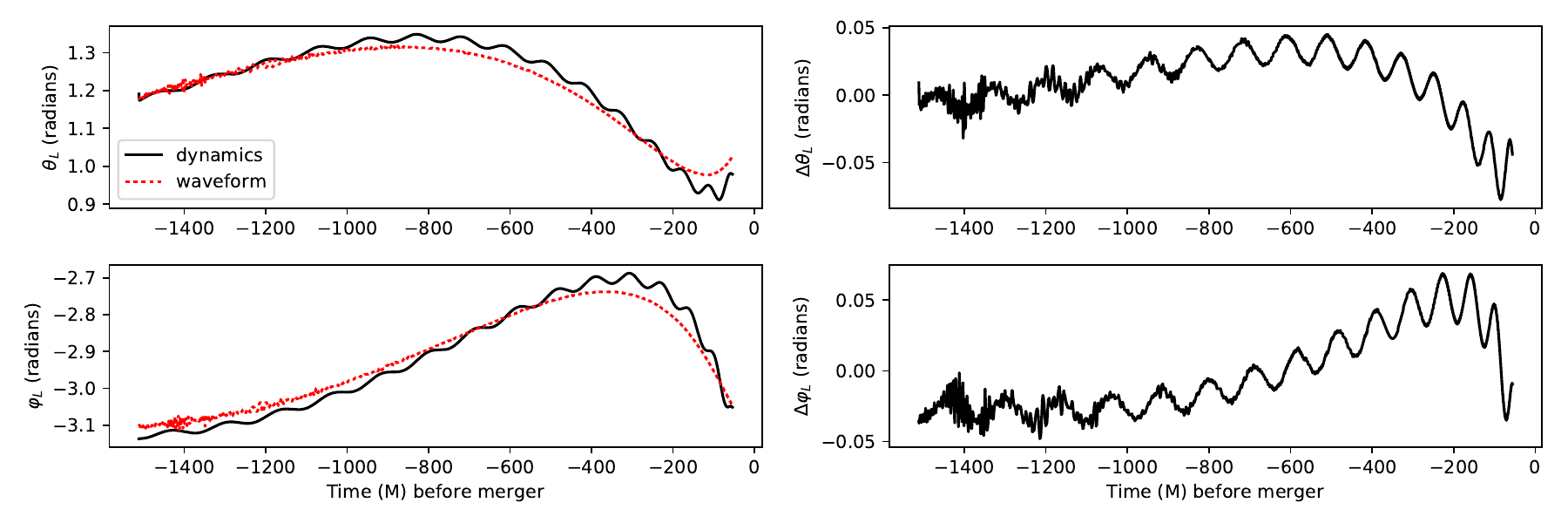} 
   \caption{BAM q1 ($q=1$, $M \omega_{22}^{\rm start} = 0.0354$, $\chi_{1} = (0,-0.2,0)$, $\chi_{2}=0$). On the left is shown the evolution of $\theta_L$ and $\varphi_L$ as calculated for the Newtonian orbital angular momentum direction from the dynamics and for the direction of maximum emission from the waveform. The right hand side shows the difference between these quantities. The time shift used for this comparison is that obtained by aligning the orbital and signal phases.}
   \label{fig: NRL_comparison}
\end{figure*}

We first consider the example of a PN waveform; the details of the method to construct this waveform are summarised in 
Ref.~\cite{Schmidt:2012rh}. The PN waveform has the advantage that there is no time shift required between the waveform and the dynamics, removing the retarded-time ambiguity. 
Our example is a mass-ratio 1:3 system, where the larger black hole has a spin of $\chi = S/m^2 = 0.75$, and the spin lies on average in the 
orbital plane. We consider a 25 000$M$-long segment of a PN waveform for this system; the orbital angular frequency range is 
$0.00491-0.00525$.

In a simple-precession configuration, the orbital angular momentum precesses around the total angular momentum, $\mathbf{J}$, and the 
precession can be described by the opening angle $\theta_L$ between the orbital and total angular momenta, and the cumulative precession angle 
$\varphi_L$. In Fig.~\ref{fig: L_comparison1} we show four calculations of $\theta_L$: the Newtonian orbital angular momentum direction, $\mathbf{\hat{L}_{N}}$
(solid black line), which exhibits nutation; the post-Newtonian angular momentum direction, $\mathbf{\hat{L}_{PN}}$ (solid red line), which precesses smoothly; 
and the QA estimates calculated from $\psi_4$ (solid green line) and $h$ (dashed blue line). From this figure we make several observations. (1) $\theta_L$
calculated from $\psi_4$ precesses smoothly, but does \emph{not} agree with the direction of $\mathbf{\hat{L}_{PN}}$. (2) $\theta_L$ calculated from $h$ 
exhibits nutation, but does not agree with the direction of $\mathbf{\hat{L}_{N}}$. We note that if we calculate the PN amplitude
using only leading-order contributions (our full PN waveform used the amplitudes from Ref.~\cite{arun2009higher}), then the QA $\theta_L$ calculated from $h$
agrees perfectly with that of $\mathbf{\hat{L}_{N}}$ (which we expect by construction), but $\theta_L$ calculated from $\psi_4$ still precesses smoothly. 
This suggests that the apparent agreement between the QA and $\mathbf{\hat{L}_{PN}}$ directions in Ref.~\cite{schmidt2011tracking} was due only to 
the use of $\psi_4$ in the QA procedure, with differences masked by gauge ambiguities, and in general these directions do not agree. Note also that
Ref.~\cite{Boyle:2014ioa} shows that the nutation in the $h$-based calculation is reduced if one includes PN signal amplitude terms that account for the
mode asymmetries that lead to out-of-plane recoil, but some nutation does remain. 

In Fig.~\ref{fig: L_comparison} we show the difference between the maximum GW emission direction $\mathbf{\hat{L}_{\psi_4}}$ as calculated from $\psi_4$, and the 
Newtonian orbital angular momentum direction, $\mathbf{\hat{L}_{N}}$, the post-Newtonian angular momentum direction, $\mathbf{\hat{L}_{PN}}$ and the maximum 
GW emission direction $\mathbf{\hat{L}_h}$. We see that although there are differences between different estimates 
$\Delta\theta_{L_{N}}$ and $\Delta\varphi_{L_{N}}$ are oscillatory while $\Delta\theta_{L_{PN}}$ and $\Delta\varphi_{L_{PN}}$ are smoothly varying. This is because $\mathbf{\hat{L}_{N}}$ shows nutation while $\mathbf{\hat{L}_{\psi_4}}$ and $\mathbf{\hat{L}_{PN}}$ do not~\cite{schmidt2011tracking}. 
Additionally, $\mathbf{\hat{L}_{h}}$ shows nutation. 
Note that similar behaviour is seen for NR simulations in Ref.~\cite{Lousto:2013wta}, which considers strain, $\psi_4$, and also the Bondi news, $N = \dot{h}$. 
We used $\psi_4$ to calculate both $\mathbf{\hat{L}}$ and $\mathbf{\hat{n}}$ in all subsequent examples. Although $\mathbf{\hat{L}_{\psi_4}}$ and $\mathbf{\hat{L}_{PN}}$ agree well, they are not equal. This may be due to differing 
PN orders in the description of the dynamics and of the waveform; whether the quantities converge with higher order PN treatments 
remains to be studied. Note that the nutation in the dynamics can be removed by using an orbit-averaged PN treatment, in which case it is
the GW-based precession that exhibits nutation~\cite{Ochsner:2012dj}, but this is not consistent with the fully general-relativistic results of
NR simulations, as the later examples will illustrate. 

We also compared the orbital phase of the waveform with the coprecessing GW phase. The orbital phase was found by integrating the orbital frequency from the PN equations 
and setting the integration constant using the method described in Sec.~\ref{sec: orbphase}. The result of this comparison is shown in Fig.~\ref{fig: PN}. As can be seen, 
they agree very well over the whole 25000M of inspiral.

Fig.~\ref{fig: NRL_comparison} shows a similar comparison for an NR simulation, the BAM q1 configuration in Tab.~\ref{tab: sims}. Here the coprecessing orbital phase is found as described in Sec.~\ref{sec: orbphase}. The quantities calculated using the dynamics are time-shifted 
assuming a flat space time (the time shift described by Eq.~\ref{eqn: coord time}). We again see that the Newtonian dynamics exhibit nutation that is not present in the maximum emission 
direction calculated from GW signal.

\section{Numerical Comparisons}
\label{sec:comparisons}

\begin{table*}[htbp]
   \centering
   \begin{tabular}{@{} lcccc @{}} 
      \toprule
      Simulation & q & $M\omega_{\text{22}}^{\rm start}$ & $\chi_1$ & $\chi_2$ \\
      \midrule
      \multicolumn{5}{c}{\bf{Non-precessing waveforms}} \\
      \midrule
      SXS 152 & 1 & 0.0297 & $\left(0,0,0.6\right)$ & $\left(0,0,0.6\right)$\\
      BAM q8 & 8 & 0.0625 & $\left(0,0,0.672\right)$ & $\left(0,0,0.0105\right)$\\
      \midrule
      \multicolumn{5}{c}{\bf{Precessing waveforms}} \\
      \midrule
      SXS 58 & 5 & 0.0316 & $\left(0.5,0,0\right)$ & $\left(0,0,0\right)$\\
      BAM q1 & 1 & 0.0354 & $\left(0.000224,-0.2,-0.000103\right)$ & $\left(0, 0, 0\right)$ \\
      GT0718 & 2.5 & 0.0628 & $\left(-0.187, 0.221, 0.526\right)$ & $\left(0.0996, 0.541, 0.239\right)$ \\
      RIT 0168 & 2 & 0.0389 & $\left(-0.438,0.716,-0.104\right)$ & $\left(0.173,-0.373,-0.301\right)$ \\
      \bottomrule
   \end{tabular}
   \caption{List of simulations used in numerical comparisons. $M \omega_{22}^{\text{start}}$ gives the GW frequency at the start of the waveform. 
   }
   \label{tab: sims}
\end{table*}

In this section we compare our GW and dynamics based calculations of the coprecessing phases for NR waveforms produced using a representative set of
current codes. This is complicated by the ambiguities that we discussed
in the previous section, but a direct comparison provides us with a general sense of how well these different estimates agree, and whether
our method gives physically reasonable results. The NR waveforms that we used are summarised in Tab.~\ref{tab: sims}; these are either 
private BAM simulations, or simulations available through the SXS, Georgia Tech, RIT and LVC-NR catalogues~\cite{mroue2013catalog, jani2016georgia, healy2017rit}.

\begin{figure*}[htbp]
   \centering
   \includegraphics[width=1.0\textwidth]{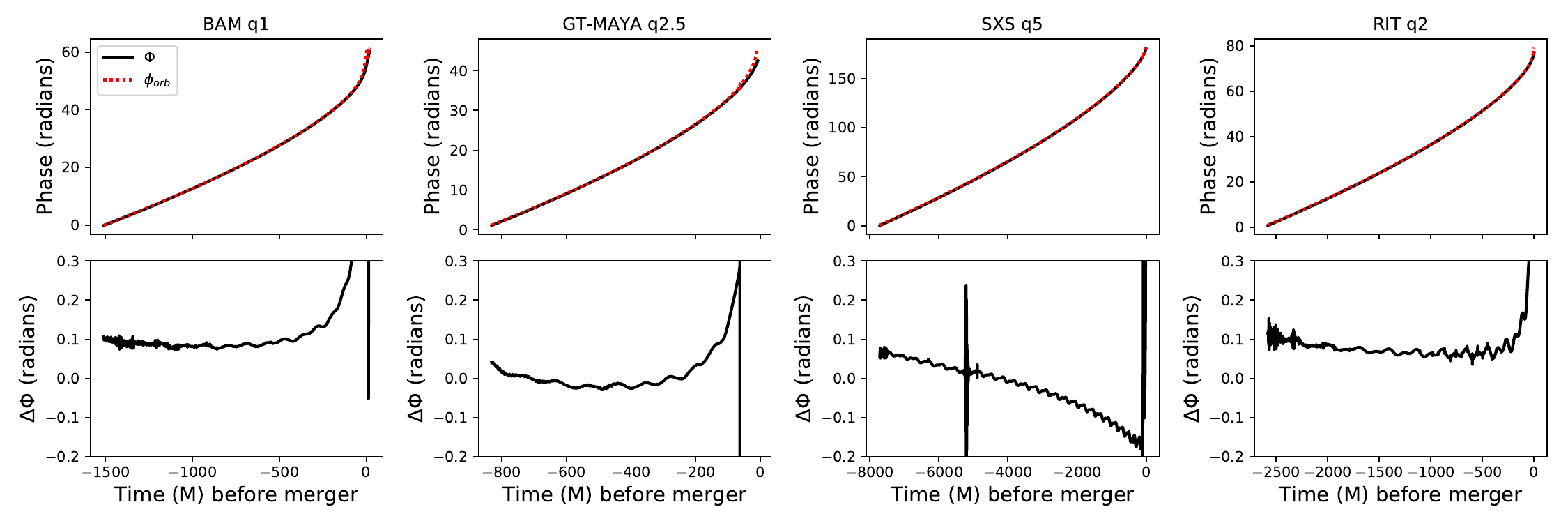} 
   \caption{A comparison of the orbital phase $\Phi$ estimated from the GW signal, and the orbital phase $\phi_{\rm orb}$ calculated from the dynamics, for the four precessing waveforms 
   listed in Tab~\ref{tab: sims}. The comparison was made using the time shifts provided by the group that produced each NR simulation.
   }
   \label{fig: 4 waveforms phase}
\end{figure*}

As for the non-precessing case, we first compared the coprecessing phases found from the waveform and the orbital motion. We chose to align the phases using the static time shift
provided with the waveform metadata, i.e., the time shift suggested by the group that produced the NR simulations. 
For the BAM, Georgia Tech and RIT waveforms, this is the value of the coordinate extraction radius, $R_{\rm ex}$. For the SpEC waveforms it is the tortoise coordinate calculated
from the areal radius; see Sec.~\ref{sec: opt ts}.
The phases agree well, as can be seen from Fig.~\ref{fig: 4 waveforms phase}. The discrepancy between the two values arises predominantly from the static time shift used to compare them. 

\begin{figure*}[htbp]
   \centering
   \includegraphics[width=1.0\textwidth]{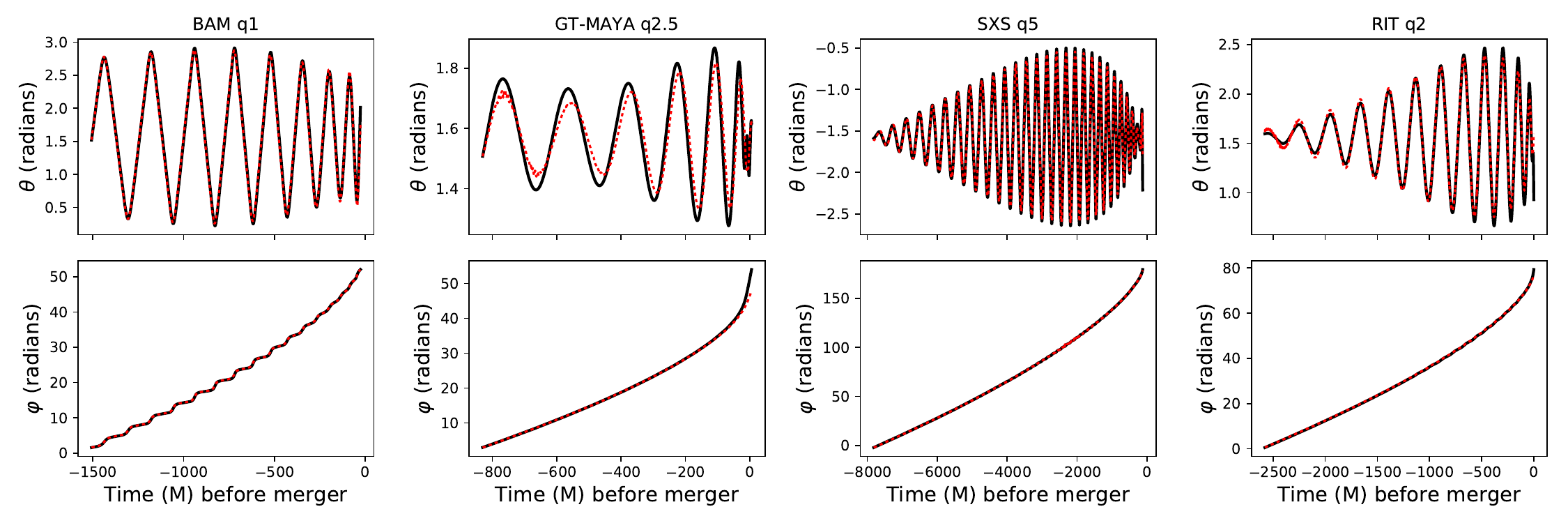} 
   \caption{A comparison of $\mathbf{\hat{n}}$ calculated from the dynamics (solid black) and using the waveform phase (dashed red) for the four precessing waveforms listed in Tab~\ref{tab: sims}. 
   This comparison was made using the time shifts provided by the group that produced each NR simulation.}
   \label{fig: 3 waveforms}
\end{figure*}

\begin{figure*}[htbp]
   \centering
   \includegraphics[width=1.0\textwidth]{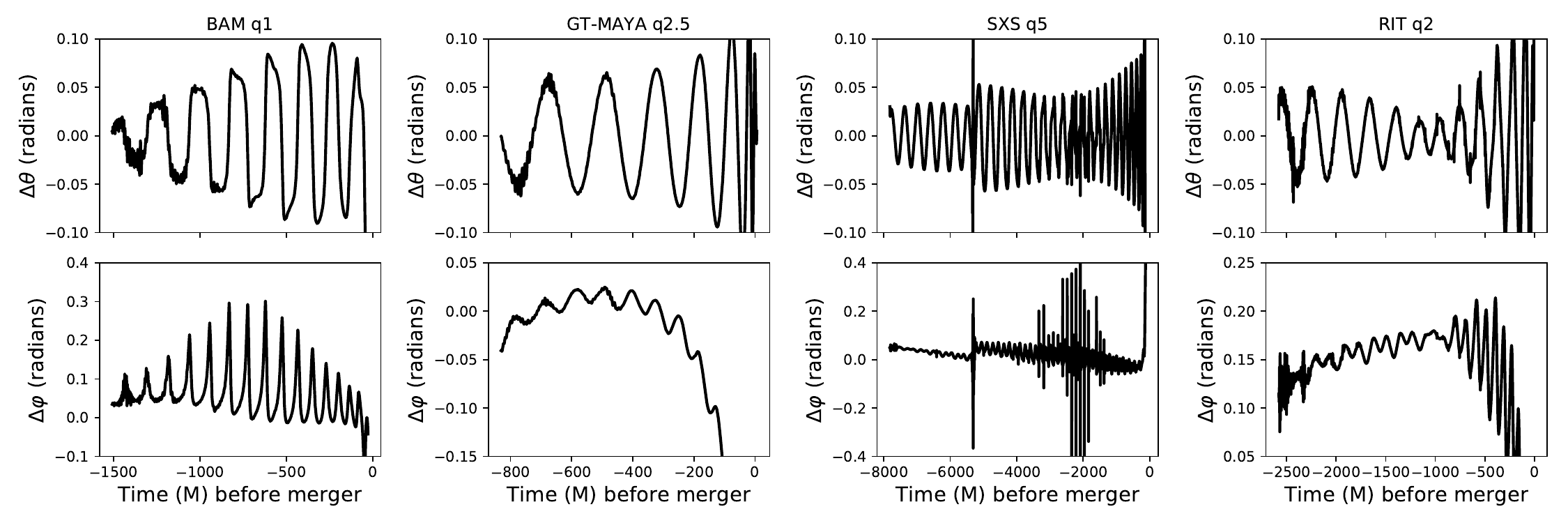} 
   \caption{A comparison of the difference between $\theta$ and $\varphi$ (shown in Fig.~\ref{fig: 3 waveforms}) calculated from the dynamics and from the waveform for the four precessing waveforms 
   listed in Tab~\ref{tab: sims}. This comparison was made using the time shifts provided by the group that produced each NR simulation.}
   \label{fig: 3 waveforms error}
\end{figure*}

We also compared the GW and dynamics based estimates 
of $\mathbf{\hat{n}}$ (i.e., $\mathbf{\hat{n}}_{w}$ and $\mathbf{\hat{n}}_{d}$) in order to observe what impact differences in phase estimates had on the quantities that are directly used by the NR Injection Infrastructure.
We calculated $\theta$, the angle between $\mathbf{\hat{n}}$ and the $z$-axis, and $\phi$, the cumulative angle between the projection of $\mathbf{\hat{n}}$ in the $xy$ plane and the $x$-axis. We then found the difference in the quantities calculated from the dynamics ($d$) and those calculated from the waveform ($w$), given by $\Delta\theta = \theta_{w} - \theta_{d}$ and $\Delta\varphi = \varphi_{w} - \varphi_{d}$.  
These are shown in Figs.~\ref{fig: 3 waveforms} and \ref{fig: 3 waveforms error}. For the BAM q1 waveform, $\mathbf{J}$ is almost along the $x$-direction, leading to large oscillations in the orientation of the
orbital plane with respect to the $z$-axis, and consequently also in $\theta$. In the other simulations 
$\mathbf{J}$ is approximately aligned in the $z$-direction, leading to smaller oscillations. We can see from the general agreement between the GW and dynamics based quantities, that our method to 
find $\mathbf{\hat{n}}$ is reliable regardless of the simulation's initial configuration. 
The SXS, RIT and GATech simulations all have $\mathbf{\hat{L}}$ approximately aligned in the $z$-direction at the beginning of the simulation. For the SXS and RIT waveform this accounts for the growth in the amplitude of the $\theta$ oscillations with time. The oscillations may not change much in amplitude for the GATech waveform because it is relatively short so may not display much of the precession cycle.

We see that our method of estimating the orbital phase $\Phi$ and unit separation $\mathbf{\hat{n}}$ using the waveform reproduces the value calculated from the dynamics
to reasonable accuracy. The levels of disagreement are consistent with the retarded-time and coordinate ambiguities, and the approximations inherent in the QA procedure.

\section{Conclusions} 
\label{sec:conc} 

We have extended previous work, which calculates a variant of the orbital angular momentum $\mathbf{\hat{L}}$ based entirely on the GW 
signal~\cite{schmidt2011tracking, o2011efficient, boyle2011geometric}, to also calculate an effective oribtal phase, $\Phi$.  These can be used
to prescribe the binary orientation and orbital phase when using NR waveforms as proxy GW signals. The most immediate
application is through the NR Injection Infrastructure used by the LIGO-Virgo collaboration~\cite{schmidt2017numerical}, and we follow the same notation and 
conventions. Our method makes it possible to orient the source without reference to the gauge-dependent binary dynamics, or a 
retarded time, which lacks a unique definition. The results of this method are in principle gauge invariant (up to finite-extraction-radius errors
in the NR waveforms), and can be used agnostically on all current binary-black-hole NR waveforms. 

As part of the validation of our method, we have compared the results to those found from the coordinate dynamics. The differences between
the two approaches are consistent with ambiguities in the definition of the retarded time, and the smoother precession of the GW-based calculation 
of precession as compared to that from the orbital dynamics. 

We note that the current NR Injection Infrastructure does not specify a choice of several conventions in the NR wave extraction (see Sec.~\ref{sec:conventions}).
In calculating the orbital phase it is necessary to take into account the choice of conventions used in extracting the NR waveforms.
 
The remaining dynamical quantities that are not considered in our method are the individual black-hole spin vectors, and the separation between 
the two black holes. The separation is not used as an observable in GW astronomy applications. Potential extensions of our method to include
the time-evolution of the spin vectors is left to future work. 

Given that our method provides a unique, gauge-invariant measure of $(\mathbf{\hat{L}}(t), \mathbf{\hat{n}}(t))$ to prescribe binary configurations, 
we recommend it as the standard measure of these quantities in the NR Injection Infrastructure. 

\section{Acknowledgements}

We thank Lionel London and Frank Ohme for useful discussions, and also Harald Pfeiffer and Patricia Schmidt for discussions of the NR Injection 
Infrastructure, and Edward Fauchon-Jones and Sebastian Khan for assistance with the LVC NR respository.
This work was supported by Science and Technology Facilities Council (STFC) grant ST/L000962/1 and European Research Council Consolidator Grant 647839.
Original BAM simulations used in this work were performed on the UK DiRAC Datacentric cluster. 

\bibliographystyle{apsrev4-1}
\bibliography{orbital_dynamics}

\end{document}